# Grid Value Analysis of Medium Voltage Back-to-Back Converter on DER Hosting Enhancement

Xiangqi Zhu, *Senior Member, IEEE*, Akanksha Singh, *Senior Member, IEEE*, Barry Mather, *Senior Member, IEEE*

*Abstract*—This paper presents an analysis of the value that can be realized by medium-voltage back-to-back (MVB2B) converters in terms of increased utilization rate of distributed energy resource (DER) and the improvement in operational conditions. A systematic, transferrable, and scalable methodology has been designed to analyze and quantify the increased DER value from three perspectives: 1) curtailment reduction of the DER generation, 2) size reduction of the energy storage needed to otherwise realize DER hosting levels, and 3) hosting capacity improvement of DER compared to base distribution circuit capability. In the case study, the proposed methodology is applied to two utility distribution systems for analysis and quantification of the grid value of the MVB2B converter, installed in the distribution circuit, and provided to the solar photovoltaic (PV) DERs. The analysis results demonstrate that the MVB2B converter can deliver significant value to PV hosting enhancement of two adjacent distribution systems when they are connected by the MVB2B converter. Based on this case study, this paper analyzes and summarizes the approximate realized grid value of the MVB2B converter for distribution systems dominated by different shares of customer classes.

## I. INTRODUCTION

DISTRIBUTED energy resources (DERs) such as photovoltaic (PV) farms have become a promising solution to transform renewable energy to satisfy the growing electricity demand [1]. The energy need has transformed this period with the disruptive change of changing power systems focused on a few centralized resources toward arcades dominated by extensive assortments of variable DERs.

For a long time, the inherent intermittency of DERs transforming renewable energy has positioned challenges for their adoptions [2]. Solutions to address those challenges have been widely studied in the literature such as demand response [3]-[5] and equipping large energy storage [6]-[7]. However, one promising solution has rarely been researched, which is power exchange among distribution feeders. If the distribution feeders can have controlled power flow exchange, the value of the DERs in the feeders are expected to be effectively enlarged by utilizing the load-DER interaction differences between the feeders.

The connections among the feeders would require the insertion of power converters or line-frequency transformers [8]-[10], whose main responsibility is to harmonize the different operating conditions between the feeders. Between the above two options, line-frequency transformers cannot control power flow which will greatly limit the flexibility and value which the connections can bring to the feeders; power converters can help feeders maintain a radio structure, and also have functions controlling power flow that can be used to develop effective solutions for problems brought by high penetrations of DERs. Therefore, adding power converters is a more universal approach that is likely to draw more interest. Also, power converters can have functions.

Now, the technology to build a medium voltage back-to-back (MVB2B) converter is becoming commercially viable, instead of staying at academic research level. Therefore, it is the time when value analysis of the MVB2B is critical to push forward the market adoption.

Currently, most research in the literature focuses on the design and control of power converters to achieve smaller size and better performance with reduced costs [10]-[12]. However, not enough work in the current state of the art analyzes questions about the benefit of implementing these converters on the grid and the value they can bring because of functions that traditional transformers do not have. Resolving these questions can help grid operators and converter vendors have a comprehensive understanding of how to properly design the functions of the MVB2B converter and effectively scheme the grid applications.

Our work in [13] investigated the value of some grid applications that can be enabled by MVB2B converters. In the initial investigation of several representative scenarios, the quantified power transfer function of the MVB2B converter demonstrated substantial benefit on most distribution circuits. To obtain a statistically meaningful quantification of the benefit and to provide effective value analysis conclusions and suggestions, this paper develops a systematic, transferrable, and scalable methodology to analyze and quantify the value that the MVB2B converter can bring to the distribution system DER hosting enhancement. The developed methodology is applied to a case study where utility-provided realistic load and PV profiles and system models are used for analysis. The analysis results provide not only a quantification of the value that the MVB2B converter can bring to various aspects but also insight into converter size selection and implementation.

The remainder of this paper is organized as follows: Section II briefly introduces the MVB2B converter. Section III presents the methodology developed for value analysis. Section IV discusses the data preparation method, and Section V presents the case study. Section VI concludes the paper and discusses future work.

## II. MVB2B Converter Introduction

This section briefly introduces the architecture of the MVB2B converter controls and the advanced grid support functions embedded in the controls.

The full controller block diagram for the MVB2B converter system is shown in Fig.1. The MVB2B converter is controlled by implementing two forms of generic switching-level control and voltage/current control [11]. The applications of the controls are in the form of an averaged current control setup, with an outer voltage loop, and an inner current loop. The voltage control is further split into three different controllers: active power, reactive power, and dc-link control.

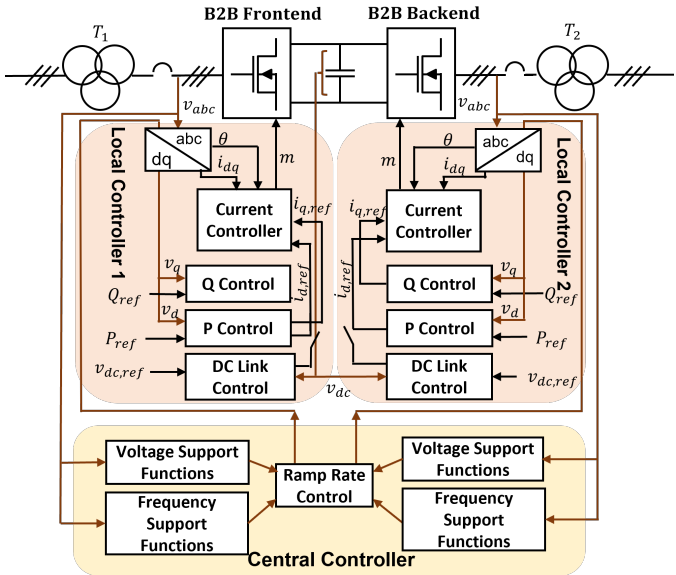

Fig. 1 MVB2B converter controller block diagram

The decentralized hierarchical controller architecture encompasses two levels of control: the central control and the local control. The central controller contains the advanced grid support functions for integration into the grid, whereas the local controllers contain the lower-level current control and switching controls. The grid measurements are used to generate the voltage reference points for the system to implement the grid support functions using the profiles set in the central controller as described in [14].

The advanced grid support functions accommodated in the central controller are listed in lower part of Fig. 1. The details of the control implementations and evaluations for the advanced grid support functions accommodated in the central controller are described in [15]. These developed controls enable various functionalities used for the value analysis presented in this paper.

## III. Analysis Methodology

This section introduces the methodology developed for the grid value analysis of the MVB2B converter on DER hosting enhancement, including the analysis metrics, the metrics quantification approach, and the preparation of a statistically meaningful data set for simulation and demonstration.

### A. Analysis Metrics

To effectively quantify the value to the grid that could be realized via the use of a MVB2B within a distribution system with high shares for DER, three analysis metrics have been designed:

1) Curtailment reduction of the DER generation
2) Size reduction of the energy storage needed for excess DER generation
3) Voltage-constrained hosting capacity improvement of the DER

As shown in (1), the curtailment reduction is measured by the reduction rate, $r_{E_c}$, which is calculated based on the difference between the curtailed DER energy before and after placing the MVB2B converter as a connection between the two feeders. The energy storage size reduction is measured by two sub-metrics: energy capacity reduction rate, $r_{E_{ES}}$, and power rating reduction rate, $r_{P_{ES}}$, as calculated in (2) and (3), respectively. The hosting capacity improvement rate, $r_{C_{DER}}$, is calculated in (4), where $C'_{DER}$ is the improved hosting capacity, and $C_{DER}$ is the hosting capacity without the MVB2B converter connected between the two feeders.

$$r_{E_c} = \frac{(E_c - E'_c)}{E_c} \tag{1}$$

$$r_{E_{ES}} = \frac{(E_{ES} - E'_{ES})}{E_{ES}} \tag{2}$$

$$r_{P_{ES}} = \frac{(P_{ES} - P'_{ES})}{P} \tag{3}$$

$$r_{C_{DER}} = \frac{(C'_{DER} - C_{DER})}{C_{DER}} \tag{4}$$

### B. Metrics Quantification Approach

As shown in (5)–(7), for a specific time period that covers $n$ time steps with a resolution of $\Delta T$, the energy storage connected with the DER is assumed to store excess generation if the load, $P_{load}(t)$, is smaller than the DER generation, $P_{DER}(t)$, and vice versa - when the DER generation is not able to power the load, energy storage will be discharged to power the load. The energy capacity and power rating of the energy storage are decided by the maximum instant energy stored in the energy storage and the maximum instant excess generation needed to be charged to the energy storage, as shown in (8)–(9). For cases where the cost-benefit balance allows some tolerance on the energy storage size, a coefficient, $\eta$, can be applied to reduce the size.

The DER generation curtailment $P_{net}^{limit}$ is calculated in (10). The assumption is that there would be a back-feeding limit, $P_{net}^{limit}$, when there is high DER penetration in a distribution system to avoid potentially jeopardizing grid stability. Therefore, if there is neither energy storage nor an MVB2B converter in the system, when the net load, $P_{net}(t)$, exceeds the limit, the excess generation will be curtailed.

When the two distribution systems are connected by the MVB2B converter, the power transfer through the MVB2B converter between the two systems is calculated in (11)–(12). When System 1 has excess DER generation and the DER generation in System 2 is not enough to cover the load, the excess power from System 1 will be transferred to System 2, and vice versa. Note that when the excess DER generation from one system is much higher than the net load in the other system, only the power amount needed to cover the load or up to the converter capacity limit, $P_{B2BC}^{limit}$, will be transferred, and the remaining excess DER generation will be back-fed to the grid or stored in the local energy storage if it exceeds the limit, $P_{net}^{limit}$.

$$P_{net}(t) = P_{load}(t) - P_{DER}(t) \tag{5}$$

$$E_{ES}(t) = E_{ES}(t-1) - P_{net}(t) \cdot \Delta T \tag{6}$$

$$P_{ES}(t) = P_{net}(t) \tag{7}$$

$$E_{ES} = \eta \cdot \max \left( E_{ES}(t) \right) \tag{8}$$

$$P_{ES} = \eta \cdot \max \left( P_{ES}(t) \right) \tag{9}$$

$$E_c = \sum_{t=1}^{n} P_{net}(t) ,$$
$$\text{where } P_{net}(t) < 0, \ [P_{net}(t)] > P_{net}^{limit} \tag{10}$$

When $P_{net}^1(t) < 0, \ P_{net}^2(t) > 0$,

$$P_{B2B}^{12}(t) = \min \left( |P_{net}^1(t)|, P_{net}^2(t), P_{B2BC}^{limit} \right) \tag{11}$$

$$P_{net}^{1\prime}(t) = P_{net}^1(t) + P_{B2B}^{12}(t) \tag{12}$$

$$P_{net}^{2\prime}(t) = P_{net}^2(t) - P_{B2B}^{12}(t) \tag{13}$$

When $P_{net}^1(t) > 0, \ P_{net}^2(t) < 0$,

$$P_{B2B}^{21}(t) = \min \left( P_{net}^1(t), |P_{net}^2(t)|, , P_{B2BC}^{limit} \right) \tag{14}$$

$$P_{net}^{1\prime}(t) = P_{net}^1(t) - P_{B2B}^{21}(t) \tag{15}$$

$$P_{net}^{2\prime}(t) = P_{net}^2(t) + P_{B2B}^{21}(t) \tag{16}$$

As shown in Fig. 3, the time-series profiles of the net loads for the two systems will be input to the energy storage model [as shown in (6)–(9)] to calculate the energy storage size when there is no MVB2B converter connecting the two systems.

Then, to investigate the benefit that the MVB2B could bring by quantifiably transferring power between the systems, the two net load profiles will be input to the MVB2B model first [as shown in (11)–(16)] to obtain the updated net load profiles, and the updated net load profiles will be input to the energy storage model to obtain the updated energy storage size.

The DER generation curtailments with and without the MVB2B connecting the systems are calculated using (10) under original net load and updated net load profiles, respectively.

Given a set of time-series load and DER generation profiles, the reduction rate of the energy storage size and DER generation curtailment can be obtained by performing the steps shown in the flowchart in Fig. 2. One set of load and DER generation profiles is not enough to draw a statistically meaningful conclusion of quantifying the benefit that the MVB2B converter can bring to the two connected systems because there are many different combinations between various weather conditions and load types, as shown in Fig. 3.

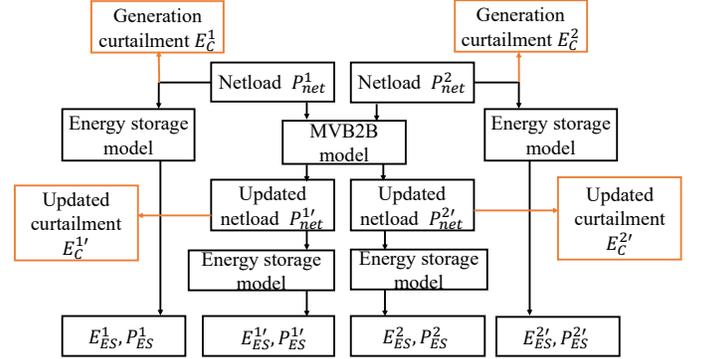

Fig. 2. Flowchart comparing energy storage size and generation curtailment

Therefore, a comprehensive database that can cover various weather and load type combinations is needed. In this paper, the Monte Carlo method is used to prepare a statistically meaningful database with limited data resources. This data preparation approach will be introduced in Section IV.

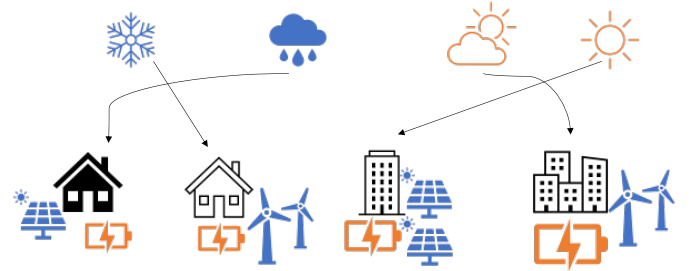

Fig. 3. Sample weather and load combinations

As shown in (17)–(20), for a database that includes N data sets with different weather and load combinations, several statistical indicators will be calculated, as shown in (17)–(20), including the maximum, average, and minimum reduction rates it can bring to the systems. Those indicators will provide a range of the reduction rate the systems can obtain after implementing the MVB2B converter. The value at a certain percentile can also be calculated to accommodate specific considerations. Equations (17)–(20) use the DER generation curtailment on System 1 as an example; the statistical indicator calculations apply to all other analysis metrics defined in (2)–(4).

$$r_{E_{Cmax}}^1 = max\left( r_{E_{C1}}^1, r_{E_{C2}}^1, \cdots, r_{E_{CN}}^1 \right) \tag{17}$$

$$r_{E_{Cmean}}^1 = mean\left( r_{E_{C1}}^1, r_{E_{C2}}^1, \cdots, r_{E_{CN}}^1 \right) \tag{18}$$

$$r_{E_{Cmin}}^1 = min\left( r_{E_{C1}}^1, r_{E_{C2}}^1, \cdots, r_{E_{CN}}^1 \right) \tag{19}$$

$$r_{E_{CX\%}}^1 = x \ percentile\left( r_{E_{C1}}^1, r_{E_{C2}}^1, \cdots, r_{E_{CN}}^1 \right) \tag{20}$$

Based on the mean value calculated in (18), the average marginal value through increasing the size of the MVB2B by $\Delta P$ will also be calculated, as shown in (21), to determine a minimum MVB2B size that can bring the maximum DER enhancement value and provide a range for the MVB2B size. This sets a foundation for MVB2B capacity sizing for real-world system implementation. A comprehensive MVB2B capacity sizing that considers the price of the converter, the price of the energy storage, the existing DER penetration level, etc., will be discussed in our follow-up paper.

$$\Delta r^1_{ECmean} = r^1_{ECmean}(P^{limit}_{B2BC} + \Delta P) - r^1_{ECmean}(P^{limit}_{B2BC}) \quad (21)$$

The DER hosting capacity improvement can be quantified by leveraging the voltage load sensitivity matrix (VLSM) developed in our previous work [16]. The voltage change that happens at bus $i$ caused by one-unit load change at bus $j$ can be quantified by the real power sensitivity factor, $p_{ij}$, and reactive power sensitivity factor, $q_{ij}$. The real/reactive power sensitivity factors constitute the VLSM for real power (VLSMP)/reactive power (VLSMQ), as shown in (22)–(24).

One common limit constraining the DER hosting capacity is the voltage limit. Large amounts of DER generation could cause a large voltage rise and cause the voltage to exceed the upper boundary.

Usually, the hosting capacity is capped by the worst scenario. Under that worst scenario, when there is an MVB2B converter connecting the two systems and connected to the bus, $\beta$, in System 1, the new voltage at a weak bus, $\alpha$, in this system can be calculated by (25) when the MVB2B converter can transfer power with an amount of $\Delta P_\beta$ to System 2. Then the capacity improvement at this bus can be calculated by (26).

Equation (27) has been derived from (25) and (26) to simplify the calculation. The calculation in (27) can be applied to multiple representative weak buses in the system, and the suggested new hosting capacity can be defined as the original capacity plus the minimum/mean $\Delta C_{DER}$ among those buses.

$$|\Delta V| = |VLSMP||\Delta P| + |VLSMQ||\Delta Q| \quad (22)$$

i.e.,

$$\begin{vmatrix} \Delta V_1 \\ \vdots \\ \Delta V_n \end{vmatrix} = \begin{vmatrix} p_{11} & \cdots & p_{1n} \\ \vdots & \ddots & \vdots \\ p_{n1} & \cdots & p_{nn} \end{vmatrix} \begin{vmatrix} \Delta P_1 \\ \vdots \\ \Delta P_n \end{vmatrix} + \begin{vmatrix} q_{11} & \cdots & q_{1n} \\ \vdots & \ddots & \vdots \\ q_{n1} & \cdots & q_{nn} \end{vmatrix} \begin{vmatrix} \Delta Q_1 \\ \vdots \\ \Delta Q_n \end{vmatrix} \quad (23)$$

derived from (2):

$$\Delta V_i = \sum_{j=1}^{n} p_{ij}\, \Delta P_j + \sum_{j=1}^{n} q_{ij}\, \Delta Q_j \quad (24)$$

$$V'_\alpha = V_\alpha + p_{\alpha\beta}\Delta P_\beta \quad (25)$$

$$\Delta C_{DER} = -\frac{V_\alpha - V'_\alpha}{p_{\alpha\alpha}} \quad (26)$$

$$\Delta C_{DER} = \frac{p_{\alpha\beta}\Delta P_\beta}{p_{\alpha\alpha}} \quad (27)$$

A case study performing these analysis metric quantification methods will be discussed in Section V.

## IV. Data Preparation

Ideally, a statistically meaningful grid value analysis and quantification of the MVB2B converter on the DER hosting enhancement needs years of load and DER generation profiles on the targeted two distribution systems. However, data resources on distribution systems are usually limited. Moreover, the load types and compositions on a distribution system could change in the future as the population moves or new commercial/industry develops.

Therefore, to create a comprehensive database that covers numerous combinations between the load and weather conditions for years, and that covers different distribution system load types and compositions to accommodate possible future changes, a Monte Carlo method is used to generate the inclusive database needed in the grid value analysis.

Based on the characteristics of the data that are accessible, first the load profiles in the data pool need to be categorized into different types (e.g., residential, commercial, industrial). Then, different data sets that have different load type compositions can be generated from the categorized pool. As shown in Fig. 4, some data sets are dominated by commercial loads, whereas others are dominated by residential loads. For data sets fully composed of residential/commercial loads, the load patterns can also be different from each other.

Given the different load data sets generated, the DER generation profiles will be shuffled to create various load-DER profile combinations, as shown in Fig. 5. The DER generation profiles can be better categorized into groups of different seasons if the DER generations are largely influenced by the weather conditions and the targeted area has distinct seasons in a year. If there are no obvious seasons, the DER profiles can be freely shuffled around a year to create different load-DER combinations.

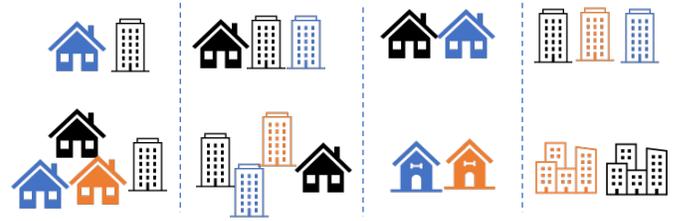

Fig.4. Sample load type combinations

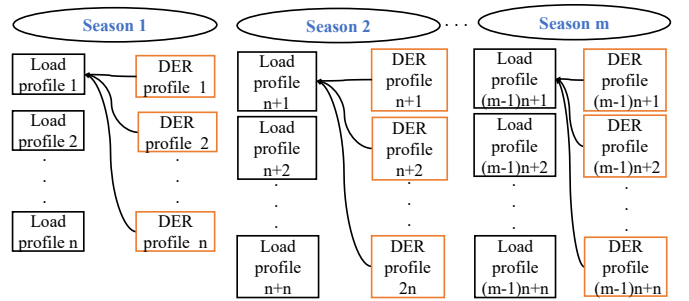

Fig. 5. Shuffle of load and DER generation profiles

## V. Case Study

This section presents the case study performed on two utility-provided realistic distribution systems and discusses the insights obtained from the results analysis. This case study is focused on the value that the MVB2B converter can bring to the solar PV hosting enhancement for the two distribution systems.

### A. Scenario Design

The two realistic distribution system models used in this analysis are shown in Fig. 6. Both systems have more than 200 load nodes. It can be observed that the connection buses for the MVB2B converter on the two systems are designated in the middle area of the systems. When selecting the system connection buses for the MVB2B converter, several factors need to be considered, such as the power transfer impact on the connection bus, the impact on other buses caused by the power change at the connection bus, and the convenience of the connection in a real-world implementation. Because this paper focuses on the value analysis, the connection bus selection will be discussed in our follow-up paper. In this case, the two buses shown in Fig. 6 are selected because the voltage impact on the connection bus and the real-world connection convenience are weighted as the most important considerations, and the connection buses could change if other considerations are placed in a more important position.

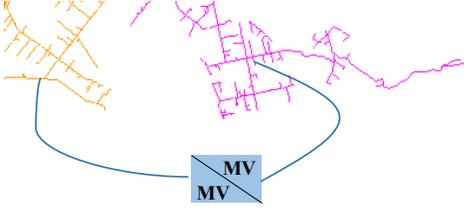

Fig. 6. Sample load and solar generation profiles in the load pool

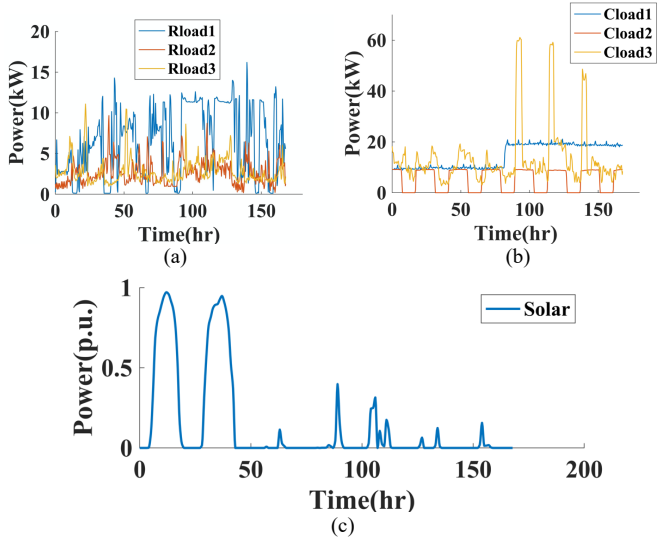

Fig. 7. Sample load and solar generation profiles in the load pool

A data pool has been built using the utility partner-provided 30-minute resolution data, including 3 years of load profile data of more than 600 commercial buildings and residential houses and 1 year of solar PV generation data. Fig. 7 shows samples of 1 week's load profiles and PV generation profiles. It can be observed that the profiles of the residential loads [Fig.7 (a)] and commercial loads [Fig. 7(b)] have obvious differences in patterns and characteristics. When combined with various PV generation profiles under different weather conditions [Fig. 7(c)], they will have various net load profiles that can create a comprehensive database for MVB2B value analysis.

By leveraging the Monte Carlo method, we built a database that includes 22 data sets, as shown in Table I and Table III.

The data sets shown in Table I are all developed under single-load scenarios, where the two systems are purely occupied by one load type. Take Set 1 as an example, the loads on System 1 are all set as residential loads, whereas the loads on System 2 are all commercial loads. Here, R represents residential, and C represents commercial.

The seven data sets are grouped into three clusters, where the two systems in one cluster have similar peak loads, and the systems in the other two clusters have different peak loads. Note that each data set has three subsets, as shown in Table II, and each subset has a different PV penetration level. Inside each subset, there are 500 year-long PV generation profiles, and each is paired with a year-long load profile to create a net load profile. So, in each set shown in Table I, there are 1,500 year-long profiles. Here, the PV penetration is defined as $PV\ capacity/_{peak\ load}$. All the profiles used in this paper are aggregated profiles at the distribution system substation, which are the summation of all the load nodes profiles selected from the data pool.

The 15 data sets in Table III are developed under mixed-load scenarios. The load profiles of the systems comprise both residential and commercial loads. The percentage of the commercial load (X%C) is defined as $C\ load\ peak/(C\ load\ peak + R\ load\ peak)$. The PV penetration levels of these 15 data sets are all 100%, which means that there are 500 net load profiles in each data set instead of 1,500, as in Table I.

TABLE I. SINGLE-LOAD SCENARIOS

| Set | System 1 | System 2 | Note |
|---|---|---|---|
| 1 | 100% R | 100% C | |
| 2 | 100% R | 100% R | Similar peak loads |
| 3 | 100% C | 100% C | |
| 4 | 100% C | 100% C | One has a slightly higher peak load. |
| 5 | 100% R | 100% R | |
| 6 | 100% C | 100% C | One has a much higher peak load. |
| 7 | 100% R | 100% R | |

TABLE II. SCENARIO DETAIL IN EACH SET OF TABLE I

| Set 1 - System 1 - 100% R | | |
|---|---|---|
| 100% PV, 500 profiles | 80% PV, 500 profiles | 50% PV, 500 profiles |

TABLE III. MIXED-LOAD SCENARIOS

| Set | System 1 | System 2 | Set | System 1 | System 2 |
|---|---|---|---|---|---|
| 8 | 20% C | 20% C | 15 | 40% C | 80% C |
| 9 | | 40% C | 16 | | 50% C |
| 10 | | 60% C | 17 | 60% C | 60% C |
| 11 | | 80% C | 18 | | 80% C |
| 12 | | 50% C | 19 | | 50% C |
| 13 | 40% C | 40% C | 20 | 80% C | 80% C |
| 14 | | 60% C | 21 | | 50% C |
| | | | 22 | 50% C | 50% C |

### B. Value Analysis

After performing the analysis methodology presented in Section III with the above data sets, the value analysis results are presented via violin plots in Figs. 8–17.

Figs. 8–12 summarize the analysis results for the seven data sets described in Table II. The violin plots provide a vision of the data distribution and mark the mean and median of the data set. The left half of the violin plot is the data distribution area plot, and the right half is duplicated from the left half to constitute a violin shape. For the x-axis labels, the one before the slash represents the scenario on System 1, and the one after the slash represents the scenario on System 2.

Each violin plot represents the results distribution of the 500 profile scenarios of a specific PV penetration level under each data set. From left to right, every three violin plots in the same color represent one set, with PV penetration from 100%, to 80%, to 50%. To demonstrate the different energy storage (shorted to "ES" in the figure titles) energy capacity reductions for System 1 and System 2 for the residential/commercial case (Set 1 in Table II), we show two figures for each system, respectively. For other metrics, we show only the results for System 1 because the data distribution and the trend of the results for System 2 are similar to that of Feeder 1, with only some differences in the mean and median values.

In addition to the metrics in (2) and (3), which are listed in Section III, the number of deep cycles is also analyzed here for energy storage. One charge/discharge cycle is defined as a deep cycle if the charge/discharge rate is greater than 80% of the energy storage power rating.

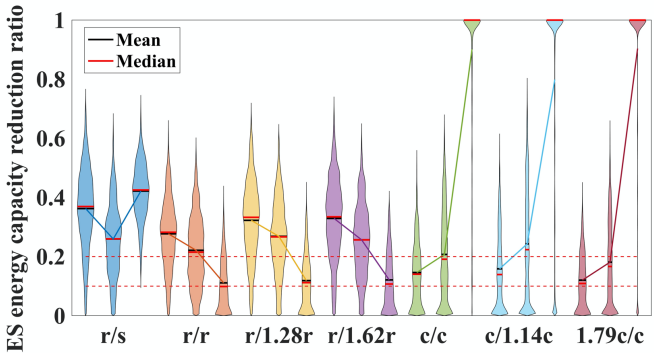

Fig. 8. Energy storage capacity reduction for data sets 1–7 on System 1

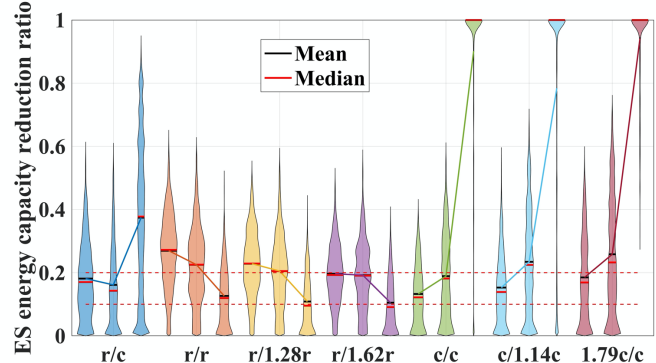

Fig. 9. Energy storage capacity reduction for data sets 1–7 on System 2

Figs. 8–9 demonstrate the energy storage capacity reduction for System 1 and System 2, respectively, over the seven data sets. If the two systems all have commercial loads, the energy capacity reduction increases when PV penetration reduces. But the trend is reversed for the scenarios where the two feeders all have residential loads. It can also be observed that the reductions brought by the converter are similar for the two systems if they both have the same types of loads. But for the case where one system is modeled with only residential loads and the other system is modeled with only commercial loads, the reduction brought to the energy storage capacity for the two

feeders is different. Figs. 8–9 shows that the residential feeder can benefit more from the converter than from the commercial feeder.

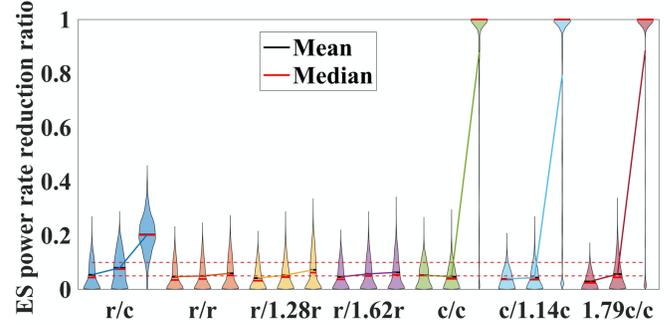

Fig. 10. Energy storage power rating reduction for data sets 1–7 on System 1

Figs. 8–10 show that the average reductions for energy storage capacity can be up to 50%, and even 100% for cases with both commercial loads and 50% PV penetration. But the reduction for power rating is not that significant - the average reductions range from 5%–10%.

It can be observed from Figs. 11–12 that the reduction in the number of deep cycles and the non-curtailed PV energy through the use of the converter are both significant—greater than 50% for most cases.

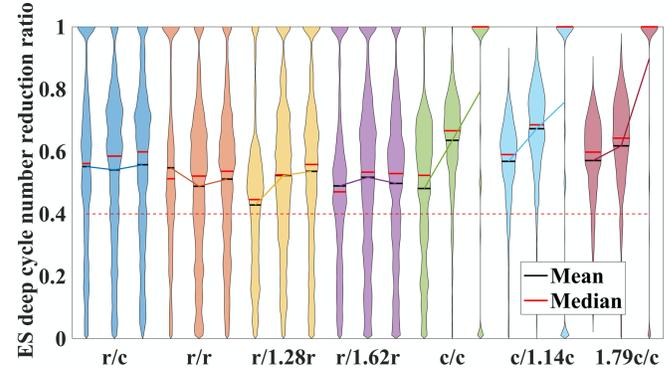

Fig. 11. Energy storage deep-cycle number reduction for data sets 1–7 on System 1

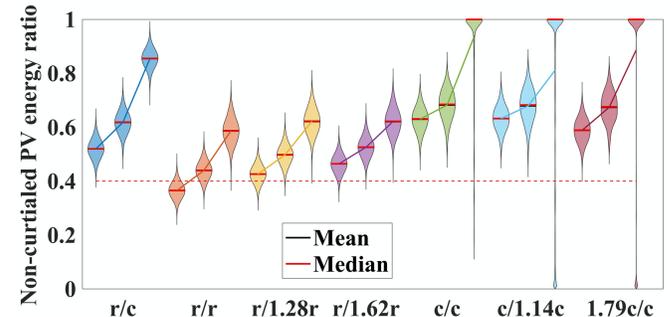

Fig. 12. Non-curtailed PV energy improvement for data sets 1–7 on System 1

Figs. 13–17 present the value analysis result for the 15 data sets described in Table IV. Similar to Figs 9–13, for the x-axis labels, the one before the slash represents the scenario on System 1, and the one after the slash represents the scenario on System 2.

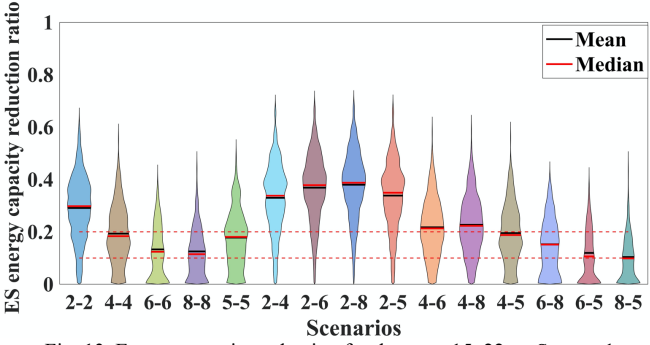

Fig. 13. Energy capacity reduction for data sets 15–22 on System 1

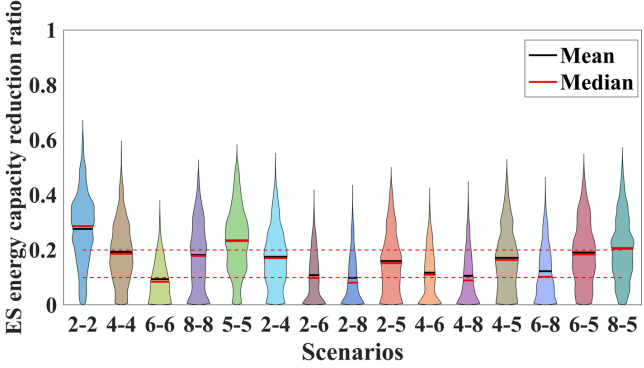

Fig. 14. Energy capacity reduction for data sets 15–22 on System 2

Comparing the violins in the right part of Fig. 13 to Fig. 14 (starting from the 6th violin from the left), it can be observed that the system with fewer commercial loads can obtain more energy storage capacity reduction when the two connected systems have an obvious difference in commercial/residential load ratio. If comparing the first four violins in Figs. 13–14, it can be observed that if the two systems have similar commercial/residential load ratios, when there are fewer commercial loads in both systems, the MVB2B converter can achieve a higher energy storage capacity reduction.

Fig. 15 shows that the power rating reduction is not very significant, ranging from 5%–10%, with some best cases achieving 20%. Many cases can achieve a reduction rate of more than 50% on deep-cycle number, as shown in Fig. 16. The PV energy curtailment reduction is also significant—greater than 50% for most cases, as shown in Fig. 17.

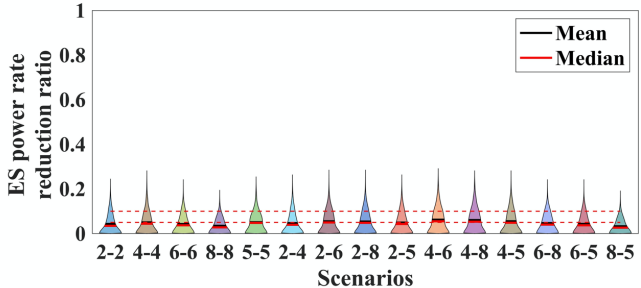

Fig. 15. Power rating reduction for data sets 15–22 on System 1

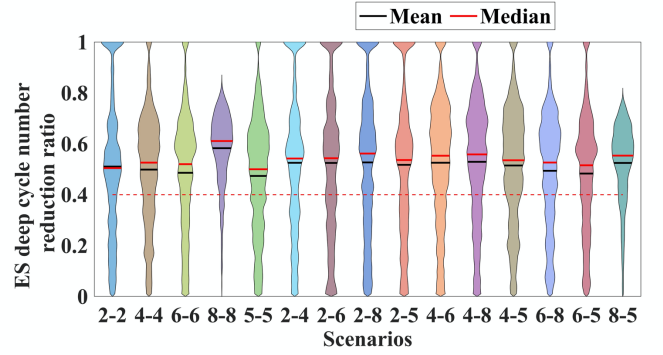

Fig. 16. System 1 Deep-cycle number reduction for data sets 15–22

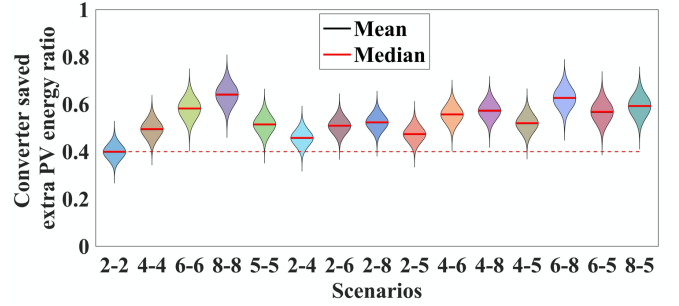

Fig. 17. System 1 Non-curtailed PV energy improvement for data sets 15–22

The PV hosting capacity improvement needs to be evaluated based on a PV hosting capacity study. In this case study, the weakest bus and worst scenario are empirically selected to demonstrate the hosting improvement calculation method described in Section III. Taking the data in Table IV and calculating the hosting capacity improvement as shown in (28) shows that the hosting improvement is approximately 20% of the maximum power the converter can transfer to another system during the worst scenario. As shown in Fig. 18, the voltage drop at this bus is not significant. If more improvement is desired, the location selection of the MVB2B converter is critical.

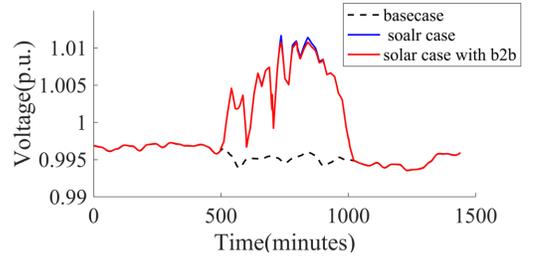

Fig. 18. One-day voltage profile of the weakest bus

TABLE IV. DATA FOR IMPROVEMENT CALCULATION

| $p_{\alpha\beta}(V/Watt)$ | $p_{\alpha\alpha}(V/Watt)$ | $\Delta P_\beta(kW)$ |
|---|---|---|
| $7.6401 \times 10^{-5}$ | $7.9190 \times 10^{-4}$ | 2000 |

$$\Delta C_{DER} = \frac{p_{\alpha\beta}\Delta P_\beta}{p_{\alpha\alpha}} = 192.9551 \text{ KW} \qquad (28)$$

### C. Marginal Value Analysis

To investigate the marginal value that the MVB2B converter can bring by increasing one unit, we performed an analysis of the non-curtailed PV energy improvement by using the converter for the 100% PV penetration subset of the seven data sets in Table. I. The converter capacity is increased from 200 kva to 750 kva, with 50 kva as a step. It can be observed that the marginal value decreases as the converter size increases, so an optimal size can be selected for each scenario.

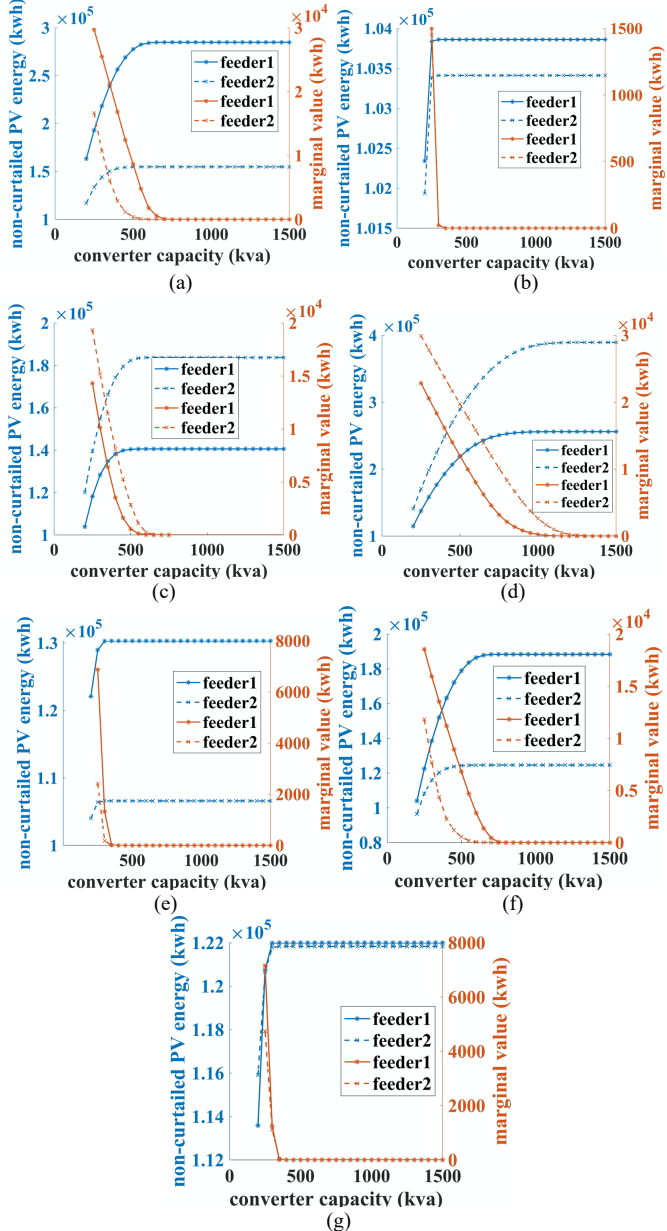

Fig. 19. Marginal value plots for non-curtailed PV energy

For the cases shown in Fig. 19 (b), Fig. 19 (e), and Fig. 19 (f), the optimal converter sizes for the two feeders are very close, so the converter size can be easily designed. For the case shown in Fig. 19 (d), continuing to increase the size of the converter should be able to lock down an optimal size. For some cases, however, such as the residential/commercial case shown in Fig. 19 (a), the optimal sizes of the two feeders are not the same. Therefore, a balance between the size of the converter and the benefit brought to the two feeders needs to be analyzed before determining the converter size. The selection of the optimal MVB2B converter size will be discussed in our follow-up paper.

## D. Insights Summary

From the case study, a couple of representative insights can be summarized:

1) The MVB2B converter can bring significant value to avoid PV curtailment in the system and to improve the PV utilization rate. If no energy storage is installed with the PV system, the MVB2B converter can save 40%–80% of the energy that might need to be curtailed if no MVB2B converter connects the two systems.

2) The MVB2B converter can improve the energy storage capacity reduction for the system with a higher portion of residential loads if the two connected systems are both mixed with residential and commercial loads. The energy storage power rating reduction is relatively small compared to the capacity reduction, whereas the deep-cycle number reduction is substantial (greater than 40% for most cases).

3) The PV hosting capacity improvement that the MVB2B converter can bring is sensitive to the location of the MVB2B connection bus.

4) The optimal size of the MVB2B converter might be different for the two connected systems if one is considering maximizing the value brought to one system. An ideal size needs to be determined by balancing the benefit brought to the two systems.

## VI. Conclusion and Future Work

This work analyzed the value that the MVB2B converter can bring to the DER hosting enhancement of distribution systems which is supported by the quantified power transfer function of the MVB2B converter. A systematic analysis methodology was proposed, and a case study on two utility-provided distribution systems was presented and discussed. The value analysis performed in the case study shows that the MVB2B converter can bring significant value to reducing PV curtailment and to reducing the number of deep cycles regardless of the load types the systems have. The energy storage size reduction is more sensitive to the load situations of the two connected systems. This also indicates that the load situations on the systems need to be considered when deciding which two systems are a better pair.

In future work, a methodology will be developed to address two problems, including selecting the optimal locations of the connection buses on the two systems and determining an optimal size of the MVB2B converter.

## Acknowledgments

This work was authored by Alliance for Sustainable Energy, LLC, the manager and operator of the National Renewable Energy Laboratory for the U.S. Department of Energy (DOE) under Contract No. DE-AC36-08GO28308. Funding provided by U.S. Department of Energy Office of Energy Efficiency and Renewable Energy Advanced Manufacturing Office via the GADTAMS project. The views expressed in the article do not necessarily represent the views of the DOE or the U.S. Government. The U.S. Government retains and the publisher, by accepting the article for publication, acknowledges that the U.S. Government retains a nonexclusive, paid-up, irrevocable, worldwide license to publish or reproduce the published form of this work, or allow others to do so, for U.S. Government purposes.